\documentclass[11pt]{emulateapj}
\usepackage{graphicx}
\usepackage{amssymb,amsmath}
\usepackage{natbib}
\begin{document}
\topmargin 0.5in 

\title{An Infrared Excess Identified in Radio-Loud Broad Absorption Line Quasars}
\author{M.A. DiPompeo\altaffilmark{1},  J.C. Runnoe\altaffilmark{1}, M.S. Brotherton\altaffilmark{1}, A.D. Myers\altaffilmark{1}}
\altaffiltext{1}{University of Wyoming, Dept. of Physics and Astronomy 3905, 1000 E. University, Laramie, WY 82071, USA}

\begin{abstract}
If broad absorption line (BAL) quasars represent a high covering fraction evolutionary state (even if this is not the sole factor governing the presence of BALs), it is expected that they should show an excess of mid-infrared radiation compared to normal quasars.  Some previous studies have suggested that this is not the case.  We perform the first analysis of the IR properties of radio-loud BAL quasars, using IR data from WISE and optical (rest-frame ultraviolet) data from SDSS, and compare the BAL quasar sample with a well-matched sample of unabsorbed quasars.  We find a statistically significant excess in the mid- to near-infrared luminosities of BAL quasars, particularly at rest-frame wavelengths of 1.5 and 4 $\mu$m.  Our sample was previously used to show that BALs are observed along many lines of sight towards quasars, but with an overabundance of more edge-on sources, suggesting that orientation factors into the appearance of BALs.  The evidence here---of a difference in IR luminosities between BAL quasars and unabsorbed quasars---can be ascribed to evolution. This suggests that a merging of the current BAL paradigms is needed to fully describe the class.

\end{abstract}


\section{INTRODUCTION}
Quasar outflows may have important effects on the evolution of quasars themselves, as well as on quasars' host galaxies and surrounding environment.  For example, AGN can theoretically affect star formation rates in the host galaxy (e.g.\ Hopkins \& Elvis 2010), and there is recent observational support of these theoretical predictions (Cano-D\'{i}az et al.\ 2012).  Properly incorporating broad absorption line (BAL) outflows into such models requires an understanding of the true nature and geometry of BALs.

BAL quasars make up an important and significant fraction of the quasar population, with observed fractions around 20\% in optically selected samples and intrinsic fractions likely higher (Hewett \& Foltz 2003, Knigge et al.\ 2008).  The debate surrounding their nature has focused on two main scenarios.  In one, BAL outflows only occur equatorially (in a loose sense, as in a true equatorial direction most quasars are likely obscured), flowing away from the accretion disk symmetry axis or the radio jet axis in radio-loud sources (Weymann et al.\ 1991, Elvis 2000).  Thus BAL winds are present in all or most quasars, and we only see them when observing at relatively large viewing angles.  In the other scenario, the BAL phase represents a short (possibly recurring) phase in the lifetime of all quasars, in which a high covering fraction cocoon of gas and dust is expelled (e.g.\ Gregg et al.\ 2006).  If this phase only has a short temporal overlap with the radio-loud quasar phase, it could exlain the dearth of radio-loud BAL quasars.

The majority of recent work has suggested that the orientation/evolution dichotomy is much too simple, and a combination of evolution and orientation is likely at work.  For example, Becker et al.\ (2000), Montenegro-Montes et al.\ (2008) and Fine et al.\ (2011) showed that BAL and non-BAL quasars have similar distributions of radio spectral index (a useful ensemble orientation indicator; see e.g.\ DiPompeo et al.\ 2012a), suggestive of similar viewing-angle ranges.  DiPompeo et al.\ (2011) extended this to a much larger and well-matched sample, with similar results but a notable over-abundance of steep-spectrum (i.e. more edge-on) BAL quasars.  Modeling these results indicated that both BAL and non-BAL quasars can be seen from small viewing angles, but beyond about 30$^{\circ}$ one is much more likely to see BALs.  There is other evidence of polar BAL outflows as well (e.g. Zhou et al.\ 2006, Ghosh \& Punsly 2007).  These results suggest that orientation plays a role in the presence of BAL features, but is not sufficient to explain the whole class.  Similar conclusions were reached by Shankar et al.\ (2008) via analysis of the BAL fraction as a function of radio power, and Allen et al.\ (2011) when looking at the BAL fraction as a function of redshift.  Finally, Bruni et al.\ (2012) find evidence that radio-loud BAL quasars are not younger than radio-loud non-BAL sources based on the relative fractions of compact steep-spectrum (CSS) radio sources in each class.

One test of high-covering-fraction evolutionary scenarios is to make detailed comparisons of the infrared properties of BAL and non-BAL sources.  It is well-known that BAL spectra are on average more reddened, especially when low-ionization BAL (LoBAL) quasars are considered (Sprayberry \& Foltz 1992, Brotherton et al.\ 2001, DiPompeo et al.\ 2012b).  However, this simply indicates the presence of significant line-of-sight dust in the wind.  Willott et al.\ (2003) found no difference in the sub-millimeter properties of BAL and non-BAL quasars, while Gallagher et al.\ (2007; henceforth G07) examined the SEDs of 38 BAL quasars and found little evidence to suggest that the mid-IR properties of BAL quasars are signficantly different from non-BALs.  Lazarova et al.\ (2012; henceforth L12) reached similar conclusions using a volume-limited sample of 22 LoBAL quasars.

In this article, we present an analysis of the IR properties of the large (73 object) radio-loud BAL quasar sample of DiPompeo et al.\ (2011) using data from the Wide-Field Infrared Survey Explorer (WISE; Wright et al.\ 2010).

\section{THE DATA}

The samples utilized in this work are those from DiPompeo et al.\ (2011), which include 73 BAL quasars from the catalog of Gibson et al.\ (2009) that have 1.4GHz radio flux densities above 10 mJy in the FIRST survey (Becker et al.\ 1995).  We also utilize the one-to-one matched sample of non-BAL quasars in DiPompeo et al.\ (2011) for comparison here; these sources are individually matched with each BAL quasar based on observed SDSS $i$-band magnitude, redshift, and 1.4GHz flux.  See DiPompeo et al.\ (2011) for full details of the samples.

We searched the WISE all-sky source catalog for matches to the optical (SDSS) coordinates within 2\arcsec\ (where the number of matches levels off before finding multiple objects per source).  In the four WISE bands (3.4, 4.6, 12, and 22 $\mu$m observed frame), we find definite matches (with SNR$>$2) for 72 (99\%), 72 (99\%), 69 (95\%), and 60 (82\%) BAL quasars, and 72 (99\%), 72 (99\%), 65 (88\%), and 40 (54\%) non-BAL quasars, respectively.  There are 3 and 12 BAL sources with 95\% confidence upper limits at 12 and 22 $\mu$m, respectively, and 7 and 32 non-BAL sources with 95\% confidence upper limits, respectively.  Three sources are not detected at all; one BAL (SDSS J151630.30-005625.5) and two non-BALs (SDSS J080415.80+300430.8 and SDSS J161948.58+382729.9).  Because the number of non-detections is small, we will not consider these objects further. 

%ADM I agree it's small number statistics, but I'd be fascinated to know of reasons why these BALs might be missing. I wonder if they've faded with time in the optical or have BAL disappearance etc.? Might be worth speculating.

We handle the sources with only upper limits in two ways, as simply leaving them out will skew the results because there are more upper limits in the non-BAL sample.  Because upper limits are found only in the 12 and 22 $\mu$m bands, all sources have at least two definite data points.  In cases where 3 or 4 data points are available, we have verified that the fluxes are well described by a simple power-law.  Therefore, when only the shorter wavelengths have significant detections, we can use a power-law fit to extrapolate out to longer wavelengths and obtain reasonable estimates of the fluxes.  This works well with the exception of two objects that appear to have inverted spectra (BAL quasar SDSS J084224.38+063116.7 and non-BAL quasar SDSS J124206.57+370138.9).  We leave these sources out of any analysis at the two longer wavelengths.  Another way to account for the upper limits is to use survival analysis, which is discussed in the next section.

After converting the WISE magnitudes to fluxes using the standard WISE zero-points, we k-correct the measurements in each band to $z=2$ (the mean and median redshift of the samples).  These correspond to rest-frame wavelengths of 1.1, 1.5, 4, and 7.3 $\mu$m.  The k-corrections use IR spectral indices measured from simple power-law fits to the available WISE data and the redshifts from SDSS.  We finally calculate monochromatic IR luminosities ($\lambda L_{\lambda} (IR)$), using a cosmology where $H_0 = 71$ km s$^{-1}$ Mpc$^{-1}$, $\Omega_M=0.27$, and $\Omega_{\Lambda} =0.73$; Komatsu et al.\ 2011).
 
Although the samples are well matched, in order to make meaningful comparisons we normalize the IR luminosities by the monochromatic ultraviolet luminosity at rest-frame 2500\AA.  These values are measured using the fits to the SDSS spectra as described in DiPompeo et al.\ (2012b).  However, as reddening can affect these measurements, we make a correction to the UV luminosity for all sources following a similar method as Gregg et al.\ (2006).  Using average values of $g-i$ as a function of redshift for quasars from Richards et al.\ (2003) and the SDSS $i$-band PSF magnitudes (which are much less affected by reddening and absorption), we estimate an unreddened $g$-band magnitude and thus a value of $A_g$.  Note that we only use positive values of $A_g$; if $A_g$ is negative, indicating that the source is bluer than the average color, we set $A_g$ to 0.  Using an SMC dust extinction curve (Gordon et al.\ 2003), we convert $A_g$ to the reddening at rest-frame 2500\AA\ for each object and use this to correct $\lambda L_{\lambda} (2500\AA)$.  We then use these reddening-corrected values to normalize each $\lambda L_{\lambda} (IR)$.  There is some evidence that BAL quasars are intrinsically more blue than non-BALs (Reichard et al.\ 2003); however this is not particularly well quantified, and so we take the conservative approach here and assume the colors of BAL and non-BAL quasars are similar.  If we do assume that the average $g-i$ color of BAL quasars is 0.1 magnitudes bluer than non-BALs, the main results do not change significantly.

There is an important caveat to using this reddening-corrected UV luminosity.  While the samples were selected to be well matched in observed, unreddened luminosity, the stronger amount of reddening in BAL quasars can cause a slight mismatching in intrinsic UV luminosity, and affect the normalized IR luminosities.  While the distributions of reddening-corrected $\lambda L_{\lambda} (UV)$ are not strikingly different upon visual inspection, the means and medians of these luminosities in the two samples are different by 0.18 dex (BALs being brighter).  This will tend to make the normalized BAL IR luminosities lower, and may weaken the results discussed below.

\section{RESULTS \& DISCUSSION}
In this section, we summarize the major results of this work, compare with related previous work (in particular that of G07), and discuss some possible complications and how they could affect the interpretation of the data.

\subsection{The $\lambda L_{\lambda}(IR) / \lambda L_{\lambda}(2500\AA)$ distributions}
\begin{figure}
\begin{minipage}{9cm}
\centering
\includegraphics[width=7cm]{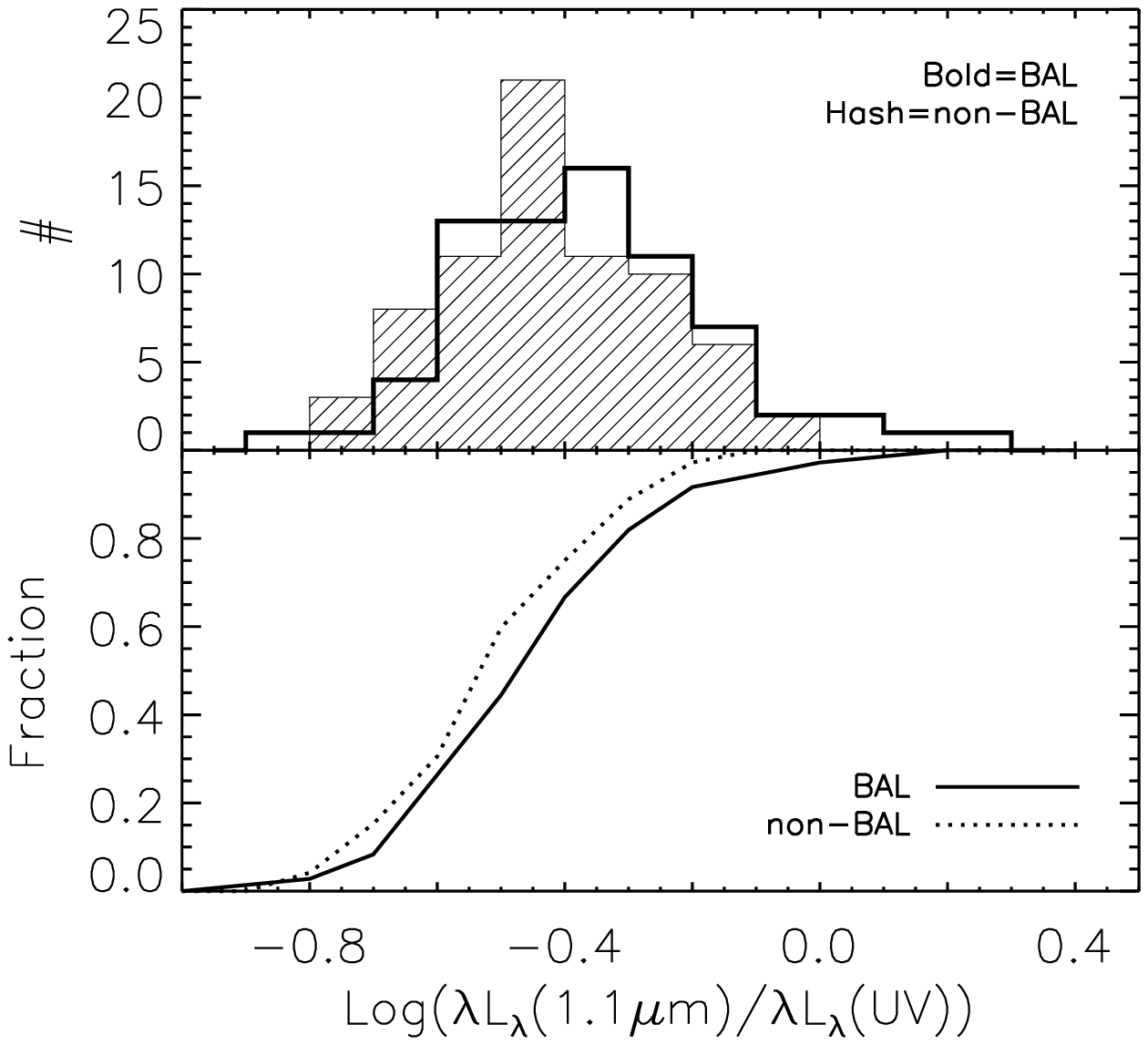}
\end{minipage}
\vspace{-0.6cm}

\begin{minipage}{9cm}
\centering
\includegraphics[width=7cm]{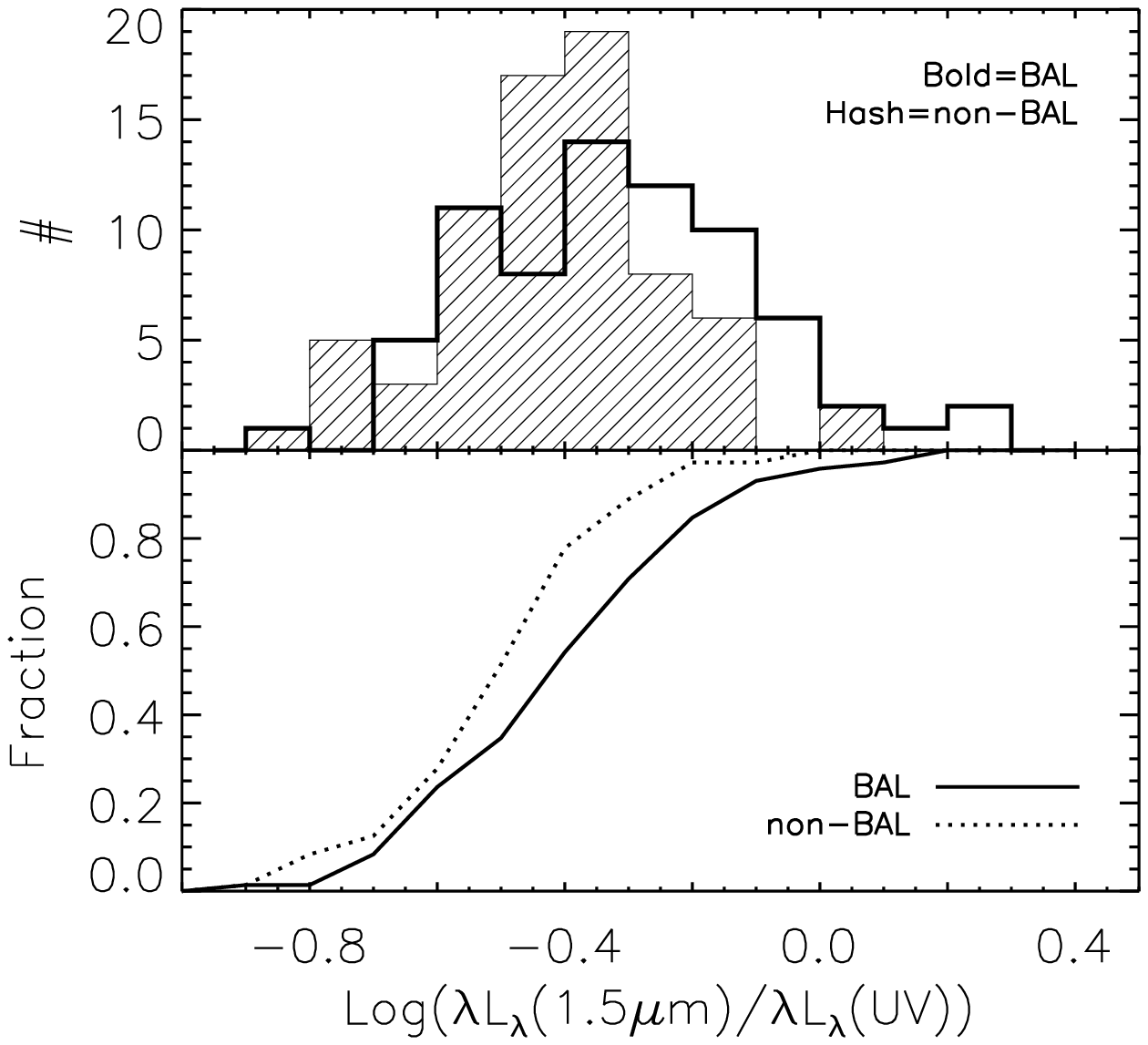}
\end{minipage}
\vspace{-0.6cm}

\begin{minipage}{9cm}
\centering
\includegraphics[width=7cm]{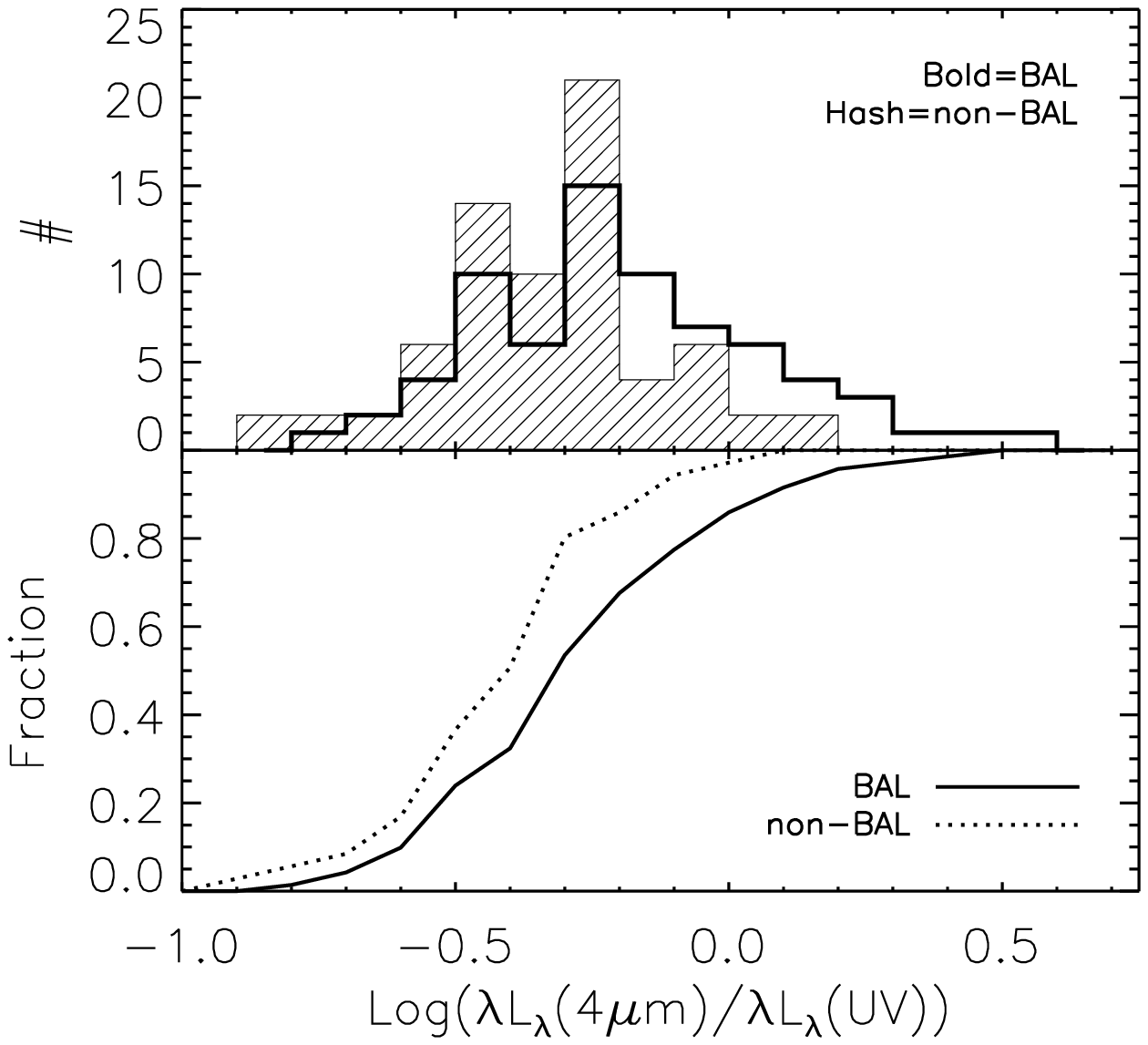}
\end{minipage}
\vspace{-0.6cm}

\begin{minipage}{9cm}
\centering
\includegraphics[width=7cm]{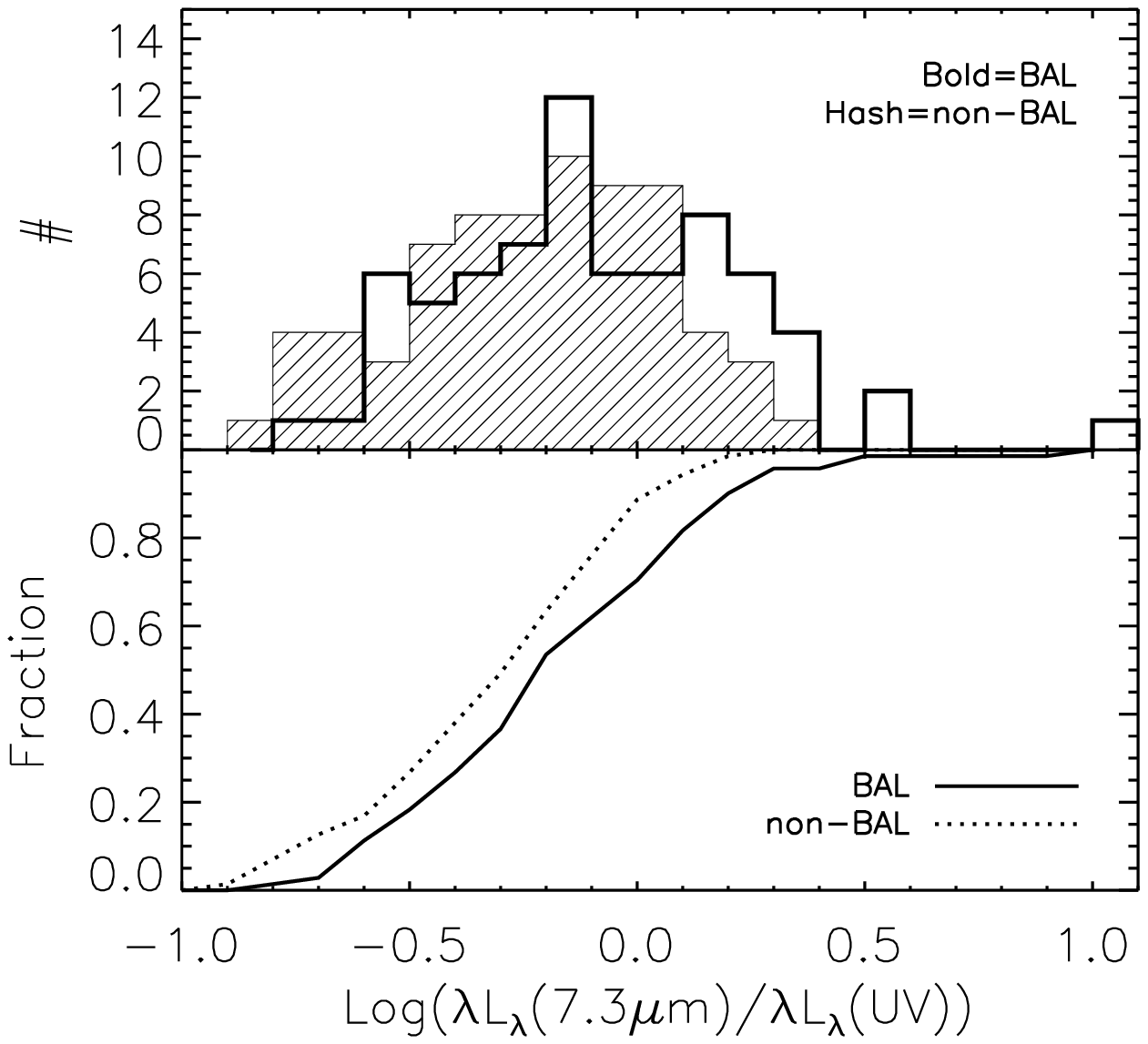}
\end{minipage}
\vspace{-0.2cm}
\caption{Comparison of IR luminosity distributions, each normalized by the reddening-corrected 2500\AA\ luminosity. The most significant differences are at 1.5 and 4 $\mu$m.\label{Lplots}}
\end{figure}

We show in Figure~\ref{Lplots} the distributions of $\lambda L_{\lambda}(IR)$/$\lambda L_{\lambda}(2500\AA)$ for each rest-frame IR wavelength, and the properties of these distributions are given in Table~\ref{basicstats}.  The normalized IR luminosity distributions at all wavelengths are wider for the BAL sample, and the means and medians are all higher, in many cases by 0.1 dex or more.  

\begin{figure}
\centering
  \figurenum{2}
   \includegraphics[width=9cm]{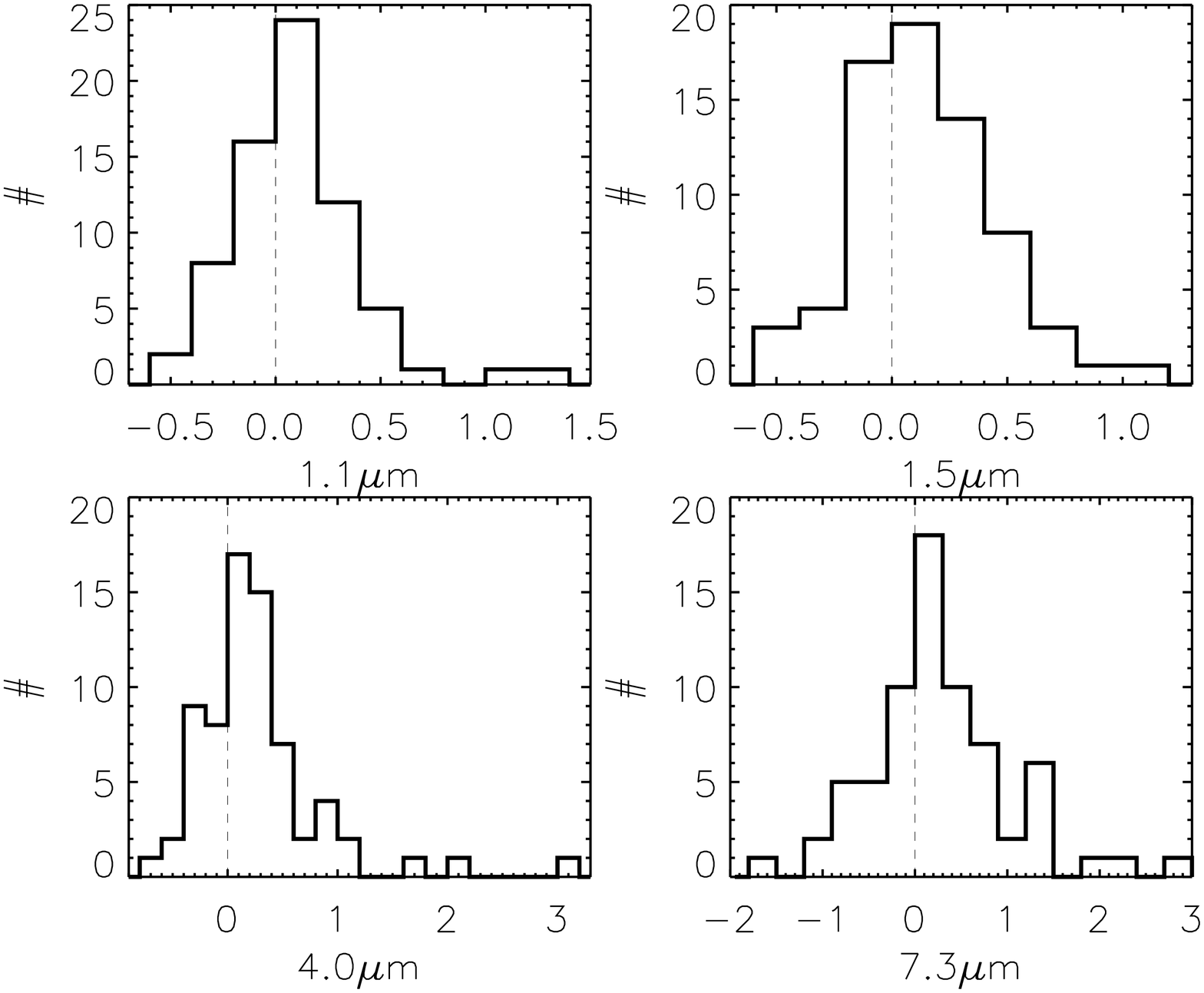}
  \caption{The distributions of the BAL minus non-BAL matched-pair IR luminosities.  It is clear that none of the distributions are centered on nor symmetric about 0.  The most significant differences are again seen at 1.5 and 4 $\mu$m (see the last two columns of Table~\ref{diststats}).\label{matchedplots}}
\end{figure}

Table~\ref{diststats} shows the results of statistical tests on the samples.  The first four rows consider all objects with the necessary data.  We perform both a Kolmogorov-Smirnov (KS) test, which tests whether the samples are drawn from the same parent population, and a Wilcoxon Rank-Sum (RS) test, which tests whether the two samples have the same means.  The results from the two tests are quite similar, and thus only the R-S test results are shown in the table for simplicity.  We will discuss the other eight rows of the table in the following sections.  In the final two columns of Table~\ref{diststats}, to the right of the vertical line,  we show the results of a matched-pair test (using a signed-rank-sum test), which tests if the distribution of the differences between each matched value is symmetric about 0.  This sample is uniquely suited for such a test.  The distributions of the BAL minus non-BAL matched pair values used for this test are shown in Figure~\ref{matchedplots}.  There are some significant outliers at the longer wavelengths in these differences, indicating that while well-matched in other properties, some pairs have very different IR luminosities.  Excluding them does not significantly change the results.

While the significance varies depending on which test is used, it is clear that there is a significant ($P<0.02$; values satisfying this cutoff are shown in bold in Table~\ref{diststats}) difference in the IR luminosities of BAL and non-BAL quasars, particularly at rest frame 1.5 and 4 $\mu$m.  The differences at 1.1 and 7.3 $\mu$m are less significant, and fall below our cutoff in the R-S test, but appear significant in the matched pair tests.

The second half of Table~\ref{diststats} shows the results without using our extrapolated fluxes, and using the WISE upper limits in a survival analysis.  We utilize the \textsc{ASURV} package in \textsc{IRAF} (LaValley et al.\ 1992) for these tests.  As expected, since the data at 1.1 and 1.5 $\mu$m have no upper limits, the results at these wavelengths are essentially identical.  And, while the significance varies, the general results discussed above still hold.

\begin{deluxetable*}{cccccccc}
 \tabletypesize{\scriptsize}
 \tablewidth{0pt}
 \tablecaption{BAL and non-BAL IR luminosity statistics\label{basicstats}}
 \tablehead{
                                 &            \multicolumn{3}{c}{all BAL}    &        &    \multicolumn{3}{c}{all non-BAL} \\
                                \cline{2-4}                                                      \cline{6-8} \\
                                & \colhead {$\mu$}  &  \colhead{Med.}  &  \colhead{$\sigma$}  & &   \colhead {$\mu$}  &  \colhead{Med.}        &  \colhead{$\sigma$} 
   }
   \startdata
   $\log(\lambda L_{\lambda}(1.1\mu m)/\lambda L_{\lambda}(UV))$  &   0.49   &   0.44    &   0.28    & &   0.41   &   0.38   &   0.16   \\
   $\log(\lambda L_{\lambda}(1.5\mu m)/\lambda L_{\lambda}(UV))$  &   0.54   &   0.48    &   0.29    & &   0.42   &   0.40   &   0.18   \\
   $\log(\lambda L_{\lambda}(4.0\mu m)/\lambda L_{\lambda}(UV))$  &   0.78    &   0.62    &   0.60    & &   0.52   &   0.46   &   0.27   \\
   $\log(\lambda L_{\lambda}(7.3\mu m)/\lambda L_{\lambda}(UV))$  &   1.10    &   0.72    &   1.29    & &   0.71   &   0.64   &   0.44  \\
   \hline \\
       &            \multicolumn{3}{c}{$BI>0$ BAL}    &        &    \multicolumn{3}{c}{non-BAL} \\
                                \cline{2-4}                                                      \cline{6-8} \\
    $\log(\lambda L_{\lambda}(1.1\mu m)/\lambda L_{\lambda}(UV))$  &   0.45   &   0.37    &   0.30    & &   0.41   &   0.38   &   0.16   \\
   $\log(\lambda L_{\lambda}(1.5\mu m)/\lambda L_{\lambda}(UV))$  &   0.47   &   0.41    &   0.27    & &   0.42   &   0.40   &   0.18   \\
   $\log(\lambda L_{\lambda}(4.0\mu m)/\lambda L_{\lambda}(UV))$  &   0.75    &   0.59    &   0.71    & &   0.52   &   0.46   &   0.27   \\
   $\log(\lambda L_{\lambda}(7.3\mu m)/\lambda L_{\lambda}(UV))$  &   1.04    &   0.67    &   1.60    & &   0.70   &   0.64   &   0.44   \\
   \hline \\
       &            \multicolumn{3}{c}{$\alpha_{rad}<-0.5$ BAL}    &        &    \multicolumn{3}{c}{$\alpha_{rad}<-0.5$ non-BAL} \\
                                \cline{2-4}                                                      \cline{6-8} \\
    $\log(\lambda L_{\lambda}(1.1\mu m)/\lambda L_{\lambda}(UV))$  &   0.52   &   0.45    &   0.31    & &   0.39   &   0.36   &   0.17   \\
   $\log(\lambda L_{\lambda}(1.5\mu m)/\lambda L_{\lambda}(UV))$  &   0.58   &   0.50    &   0.30    & &   0.39   &   0.34   &   0.20   \\
   $\log(\lambda L_{\lambda}(4.0\mu m)/\lambda L_{\lambda}(UV))$  &   0.85    &   0.63    &   0.66    & &   0.45   &   0.37   &   0.28   \\
   $\log(\lambda L_{\lambda}(7.3\mu m)/\lambda L_{\lambda}(UV))$  &   1.25    &   0.82    &   1.49    & &   0.64   &   0.48   &   0.45  
   \enddata
\end{deluxetable*}

\begin{deluxetable*}{llccccc|cc}
 \tabletypesize{\scriptsize}
 \tablewidth{0pt}
 \tablecaption{BAL and non-BAL distribution statistics\label{diststats}}
 \tablehead{
   \colhead{Sample 1} & \colhead{Sample 2}  &   \colhead{$\lambda$ ($\mu$m)}   &    \colhead{$n_1$} & \colhead{$n_2$} &  \colhead{$Z$}  & \colhead{$P_{RS}$}  & \colhead{$n_m$}  &  \colhead{$P_W$}
   }
   \startdata
Full BAL       & Full nBAL   &  1.1   &  72    &       72    &      -1.74   &  0.08    & 70  &  \textbf{0.016}     \\
Full BAL       & Full nBAL   &  1.5   &   72   &       72   &       -2.61   &   \textbf{0.008}  & 70  &  \textbf{0.0006} \\
Full BAL       & Full nBAL   &  4      &   71    &      71   &       -3.15  &   \textbf{0.001}  & 70  & \textbf{0.00007}   \\
Full BAL       & Full nBAL   &  7.3   &   71   &       71  &        -2.21  &  0.027   & 33  & \textbf{0.002}    \\
\hline
$BI>0$ BAL & Full nBAL   &  1.1   &   39   &      72   &       -0.07 &  0.946    &  37 &  0.437    \\
$BI>0$ BAL & Full nBAL   &  1.5   &   39   &      72   &      -0.52   &  0.604  &  37  & 0.214     \\
$BI>0$ BAL & Full nBAL   &  4      &   38   &      71    &      -1.51   &  0.13   & 37  &  \textbf{0.015}     \\
$BI>0$ BAL & Full nBAL   &  7.3   &   38   &      71    &      -0.61   & 0.540    & 37  &  0.031    \\
\hline
$\alpha_{rad}<-0.5$ BAL & $\alpha_{rad}<-0.5$ nBAL &  1.1   &   49   &     27   &    -2.40      & \textbf{0.016}    & 22  & \textbf{0.004}     \\
$\alpha_{rad}<-0.5$ BAL & $\alpha_{rad}<-0.5$ nBAL &  1.5   &   49   &     27   &    -3.38      & \textbf{0.0008}  & 22  & \textbf{0.001}     \\
$\alpha_{rad}<-0.5$ BAL & $\alpha_{rad}<-0.5$ nBAL &  4      &   49   &     26   &    -3.43      & \textbf{0.0003}  & 22  & \textbf{0.0003}   \\
$\alpha_{rad}<-0.5$ BAL & $\alpha_{rad}<-0.5$ nBAL &  7.3   &   49   &    26    &    -2.73      & \textbf{0.003}    & 22  & \textbf{0.004}    \\
\\
                                              &                                                 &                \multicolumn{4}{c}{Survival Tests}              &        \\
                                                                              \cline{3-6}                                                      \\
Full BAL       & Full nBAL   &  1.1   &  72   &   72         &    -1.74    &   0.08   & \nodata   & \nodata  \\ 
Full BAL       & Full nBAL   &  1.5   &  72   &   71        &     -2.61   &   \textbf{0.009}  &  \nodata  &  \nodata  \\
Full BAL       & Full nBAL   &  4      &  72    &  72        &      -3.22   &  \textbf{0.001}   & \nodata    & \nodata \\
Full BAL       & Full nBAL   &  7.3   &   72   &  71        &      -2.49   & \textbf{0.012}    & \nodata    &   \nodata\\
\hline
$BI>0$ BAL & Full nBAL   &  1.1   &   39   &      72   &       -0.07 &  0.945    &  \nodata &  \nodata    \\
$BI>0$ BAL & Full nBAL   &  1.5   &   39   &      72   &      -0.52   &  0.604  &  \nodata  & \nodata     \\
$BI>0$ BAL & Full nBAL   &  4      &   39   &      72    &      -0.98   &  0.328   & \nodata  &  \nodata     \\
$BI>0$ BAL & Full nBAL   &  7.3   &   39   &      72    &     -0.39   & 0.699    & \nodata  &  \nodata    \\
\hline
$\alpha_{rad}<-0.5$ BAL & $\alpha_{rad}<-0.5$ nBAL &  1.1   &   51   &     29        &  -2.43      & \textbf{0.015}    & \nodata  & \nodata     \\
$\alpha_{rad}<-0.5$ BAL & $\alpha_{rad}<-0.5$ nBAL &  1.5   &   51   &     29        &  -3.19      & \textbf{0.001}  & \nodata  & \nodata     \\
$\alpha_{rad}<-0.5$ BAL & $\alpha_{rad}<-0.5$ nBAL &  4      &   51   &     29        &   -3.49      & \textbf{0.0005}  & \nodata  & \nodata   \\
$\alpha_{rad}<-0.5$ BAL & $\alpha_{rad}<-0.5$ nBAL &  7.3   &   51   &    29        &  -1.95       & 0.051                  & \nodata  & \nodata    
\enddata
 \tablecomments{The top half shows the distribution test statistics including our extrapolated values of the IR luminosity at 4 and 7.3 $\mu$m (when needed).  $\lambda$ indicates the rest-frame wavelength of the k-corrected infrared luminosity.  $Z$ is the R-S test statistic and $P_{RS}$ is the corresponding probability that the BAL and non-BAL samples have the same mean.  To the right of the vertical line are the matched-pair (signed Wilcoxon rank-sum) test results; $n_m$ are the number of pairs included and $P_{W}$ is the probability that the distribution of differences is symmetric about 0.  The second half of the table shows results using survival analysis on the data simply keeping the upper limits reported by WISE.  Results from K-S and log-rank tests are not shown here, but are similar. We consider $P<0.02$ to indicate that the samples are significantly different, and present these values in bold.}
\end{deluxetable*}

\subsection{Covering fractions}
The simplest interpretation of the above results is that there is a difference in dust covering fraction between BAL and non-BAL quasars, as predicted by evolutionary models.  Limits on the covering fraction can be estimated using the ratio of infrared to bolometric (accretion disk) luminosity.  We use Equation 10 of Calderone et al.\ (2012) to determine upper and lower limits, and average the two to estimate the covering fractions.  $L_{bol}$ is determined using the reddening-corrected $\lambda L_{\lambda} (2500\AA)$ luminosity and the same bolometric correction used in DiPompeo et al.\ (2012b), and we use the IR luminosity at 4$\mu$m.  We find an average covering fraction in the BAL sample of 0.44, and 0.36 in the non-BAL sample.  These are consistent with other studies of luminous, high-z quasars (e.g.\ Maiolino et al.\ 2007).  The results here suggest a difference in covering fraction between BAL and non-BAL sources of about 10\%.

This difference in covering faction may be indicative of an evolutionary difference between BAL and non-BAL sources.  However, there are other possibilities.  For example, if all quasars have winds and a range of covering fractions, we would expect to find BALs preferentially in high-covering fraction objects simply because there are more possible lines of sight through the wind (e.g. Hamann et al.\ 1993).  Considering that the difference in covering fractions found here is relatively small, the difference in IR luminosity may simply reflect this effect.

\subsection{Comparison to previous work}
If others have searched for this effect and it has not been seen, why is it seen in this sample?  Again, we point out that this sample is selected to be particularly well matched, and that it is larger than others (G07 had 38 objects, L12 had 22).  This is also the only radio-loud BAL quasar sample analyzed in this way, and could point to a difference between radio-loud and radio-quiet objects.  

In order to test if the result is simply due to increased numbers, we perform a random sampling of our objects to the sample sizes of G07 and L12 to see how often we might expect to see the differences at the significance measured here.  Randomly sampling 38 BAL quasars and comparing to the full non-BAL sample we would only expect to see these differences about 20\% of the time.  Sampling 22 BAL quasars this drops to only around 10\%.

Another issue could be the way in which the samples are defined.  G07 use the strict definition of BAL, where the traditional $BI$ (Balnicity Index; Weymann et al.\ 1991) must be greater than 0.  L12 use a slightly modified version of $BI$ more in line with the methods of the Gibson et al.\ (2009) catalog (from which our sample is drawn), where $BI$ integration begins at 0 km/s (as opposed to 2000 km/s) and is only required to remain continuous over 1000 km/s (again as opposed to 2000 km/s in the traditional definition).  In the middle four rows of Tables~\ref{basicstats} and \ref{diststats} we show the results if we only include objects with a traditional $BI>0$ (see DiPompeo et al.\ 2012b for discussion of our $BI$ measurements).  Note that the sample size is decreased to a similar size as G07.  When only this subset is considered, the significance of the difference drops considerably (it is interesting to note however that the difference at 4 and 7.3 $\mu$m actually increases in the matched-pair test).  However, it is difficult to say whether this reduction in significance is due to the sample selection or simply because of the decrease in numbers.  $BI$ does not appear to correlate in any way to the IR properties, which seems to suggest that the difference is reduced because there are fewer objects.

Finally, the true best comparison to the G07 results is our measurement at 7.3 $\mu$m, as they correct their IR luminosities to a rest-frame wavelength of 8 $\mu$m.  They also normalize by the optical luminosity at 5000\AA\ as opposed to the UV normalization at 2500\AA\ used here; however, we feel that our reddening correction to this luminosity is robust and thus should be similar to the normalization at 5000\AA.  Additionally, there tends to be more intrinsic scatter in the 5000\AA\ luminosity compared to that at shorter wavelengths (e.g.\ Runnoe et al.\ 2012), which could introduce more scatter into the normalized IR luminosities.  G07 find a $P_{RS}$ of about 16\%, while we find $P_{RS}$ of 2.7\% (5.1\% in the survival analysis).  These are not necessarily inconsistent, and could be due to the difference in the samples.

\subsection{Other caveats}
\begin{figure}
\centering
  \figurenum{3}
   \includegraphics[width=3in]{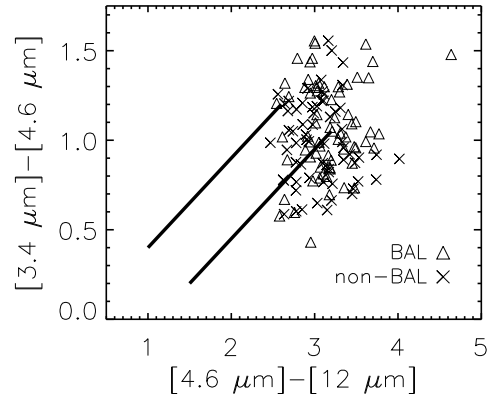}
  \caption{Our objects in WISE color-color space.  The bold lines indicate the ``blazar strip'' of Massaro et al.\ (2011); while there is some overlap between this strip and the location of quasars, none of our objects are clearly separated with an obvious blazar component.  BAL and non-BALs also generally occupy the same color space.\label{blazarstrip}}
\end{figure}

Another concern may be the average orientation of the two samples.  As noted earlier, DiPompeo et al.\ (2011, 2012a) showed that the difference in radio spectral index distributions in these samples could be explained by allowing BAL quasar viewing angles to extend about 10$^{\circ}$ farther from the symmetry axis compared to non-BAL quasars.  It is known that the optical/UV disk emission in quasars is probably anisotropic (e.g.\ Nemmen \& Brotherton 2010, Runnoe et al.\ 2012), and the IR radiation may be as well (e.g. Nenkova et al.\ 2008).  Another concern due to orientation could be synchrotron emission contributing more to the IR luminosity in more face-on sources, where relativistic beaming could boost the observed flux.  None of our objects fall on the WISE ``blazar strip'' defined by Massaro et al.\ (2011), so we do not expect this to be an important effect (Figure~\ref{blazarstrip}), but it is still worth considering.

In order to account for different average viewing angles to the two samples and address the concerns above, we consider only the subset of objects with steep radio spectra (see DiPompeo et al.\ 2011 for these measurements), which are assumed to be seen at larger angles to the jet axis.  In these subsamples, the significance of the effect is either the same or notably stronger in most cases, despite the decrease in the number of objects (Tables~\ref{basicstats} and \ref{diststats}, lower third).  If we further limit to $\alpha_{rad}>-1.5$ (the lowest value in the non-BAL sample), to best match the viewing angle ranges, the results are the same.

Some of the possible effects of orientation could actually weaken the results found here.  For example, given the average orientations from radio spectral index modeling, we would expect a larger synchrotron component in the non-BAL sample, making the distributions of IR luminosity artificially more similar than they are intrinsically.  However, all differences point to BAL quasars being systematically brighter at all IR wavelengths.

Another consideration is other contributions to the IR luminosity that are unrelated to the quasar, particularly due to star formation in the host.  If, for example, star formation rates are affected by BAL outflows, one might expect this to cause a difference in IR luminosity.  However, at the rest-frame frequencies considered here (below 10 $\mu$m), the AGN component is likely the dominant contributor (e.g.\ the $HeRG\acute{E}$ project, Drouart et al.\ 2012), and studies in the far-IR/sub-mm do not find a measurable difference in the star-formation rates of BAL and non-BAL quasars (Willott et al.\ 2003, G07, L12).

A final caveat is that the rest wavelengths probed in this study are relatively short, and generally dominated by hot dust emission.  Luminous quasars typically show bumps around 3$\mu$m, likely from dust near the sublimation temperature at the inner edge of the torus (Deo et al.\ 2011).  Orientation can then play a role in this as well, as different viewing angles will allow different lines of sight to the inner torus edge.  However, the tests on the samples limited by radio spectral index discussed above should limit this effect.  Ideally we would like to probe slightly longer wavelengths to better estimate the covering fractions, but we see no reason to believe that the results here will change significantly.  Also, as mentioned above, these wavelengths have the benefit of not being contaminated by star formation in the host galaxy, so there is some trade-off.

\section{SUMMARY \& CONCLUSIONS}
We identify a significant difference in the UV-normalized WISE-IR luminosities between radio-loud BAL and non-BAL quasars, in particular at rest-frame wavelengths of 1.5 and 4 $\mu$m.  The two samples are well-matched, and matched-pair tests suggest the differences are real.  The observed differences can be explained by an approximately 10\% greater covering fraction in BAL quasars compared to non-BALs.  When only BAL sources with $BI>0$ are considered, the differences are much less apparent, but this is likely due to a simple reduction in the number of sources as statistical tests show that the large sample size is needed to see the effect consistently.  However, it is still a possibility that this points to a real physical difference between $BI>0$ and $BI=0$ populations.  When only steep radio spectrum sources are considered, in order to avoid issues caused by average orientations of the samples, many of the differences become more significant.

These differences are difficult to reconcile with an orientation-only explanation of BAL quasars, and support some of the hypotheses of evolutionary pictures, though there are other explanations.  However, it has already been shown \textit{in this sample} that orientation likely plays a role in the presence of BALs, and the BAL and non-BAL samples have slightly different average viewing angles.  Combined with these results, the idea that some combination of orientation and evolutionary effects is at work in the BAL subclass seems more likely than ever.

\acknowledgements
MAD and ADM were supported by NASA ADAP grants NNX12AI49G and NNX12AE38G and by the National Science Foundation under grant number 1211112.  This publication makes use of data products from the Wide-field Infrared Survey Explorer, which is a joint project of the University of California, Los Angeles, and the Jet Propulsion Laboratory/California Institute of Technology, funded by the National Aeronautics and Space Administration.

\clearpage

\end{document}